# Statistical Analysis based Hypothesis Testing Method in Biological Knowledge Discovery

Md. Naseef-Ur-Rahman Chowdhury, Suvankar Paul, and Kazi Zakia Sultana

Department of Computer Science & Engineering,
Chittagong University of Engineering & Technology (CUET),
Chittagong, Bangladesh

*Abstract*

*The correlation and interactions among different biological entities comprise the biological system. Although already revealed interactions contribute to the understanding of different existing systems, researchers face many questions everyday regarding inter-relationships among entities. Their queries have potential role in exploring new relations which may open up a new area of investigation. In this paper, we introduce a text mining based method for answering the biological queries in terms of statistical computation such that researchers can come up with new knowledge discovery. It facilitates user to submit their query in natural linguistic form which can be treated as hypothesis. Our proposed approach analyzes the hypothesis and measures the p-value of the hypothesis with respect to the existing literature. Based on the measured value, the system either accepts or rejects the hypothesis from statistical point of view. Moreover, even it does not find any direct relationship among the entities of the hypothesis, it presents a network to give an integral overview of all the entities through which the entities might be related. This is also congenial for the researchers to widen their view and thus think of new hypothesis for further investigation. It assists researcher to get a quantitative evaluation of their assumptions such that they can reach a logical conclusion and thus aids in relevant re-searches of biological knowledge discovery. The system also provides the researchers a graphical interactive interface to submit their hypothesis for assessment in a more convenient way.*

## 1 Introduction

The relationships among biological entities such as proteins, genes, diseases, DNAs constitute the significant part of different biological processes. Under-standing these processes areimportant for analyzing their effects on human body and finding ways to get rid of any behavioral disorder of the systems as well. As for example, protein - protein interactions play crucial role in forming different biological systems and, therefore, any interruption can cause disaster in systems which is expressed as diseases. Different types of drugs are discovered for preventing these diseases and also for recovery from them. So study of protein interactions paves the way to comprehend the causes of diseases and discover drugs to this end. Although some recent works concentrate on this study, theymostly focus on building the network evolving the centered biological entity. But in our study, We use language processing technique to extract all direct and indirect relationships among the entities of the hypothesis and then we try to answer the query statistically before building the supporting network.

Our method is based on text mining approach. We know that the volume of literatures is growing rapidly with the increase of researches around the world. This large volume of publications holds a lot of information directly or indirectly which can be an important source for secondary information generation or for accepting or rejecting any hypothesis under experiment. If we can properly utilize this collection for the convenience of future researches, it will obviously be a potential source of knowledge. Therefore, our method stands on the basis of mining the huge volume of publications and we attempt to answer the query of the user by computing the probability of its happening with respect to its direct or indirect presence in the existing literatures. Earlier works mine the literature for exploring relationship among the entities or for building a network based on user query to broaden the prospect of the research in the area. But some cases may be raised for examining any assumption of the researchers or for reaching a conclusion about any fact for further investigation. Although many text mining based approaches have been evolved to reveal the hidden relationships among the entities, they basically stand on the co-occurrence of the terms and also do not provide an interface where a researcher can easily test a hypothesis or can statistically measure the acceptance probability of the hypothesis. In our approach, we go forward to prove or disprove user query hypothesis statistically after doing some natural language processing which will improve the accuracy of the result. Natural Language Processing also helps to curate unexplored relationships precisely. Moreover, for further research convenience, we generate a network based on the relationships with the neighboring entities which co-occur in the same literature. In this paper, we present an approach to overcome the gap in existing tools and attempt to provide a newer way to the hypothesis oriented researches.

- The tool applies natural language processing to discover the semantic meaning of the texts and follows statistical approach to compute the probability of accepting or rejecting the hypothesis and then presents it to the user such that he can reach a conclusion about the assumption.

- The tool builds a network centering the entities and their relationship directly mentioned in the hypothesis and consisting of all other entities directly or indirectly related to them. This network will facilitate the user to get an idea about other related entities and thus come up with a new area of re-search for the user.

- We give an interactive interface to the user such that he can either put his hypothesis in textual format or build his hypothesis graphically he is interested to test.

## 2 Motivation

Text mining approach is distinct from information retrieval process in the sense that information retrieval process discovers the facts that are directly published in existing literatures or already known to the respective community, on the contrary, the main goal of text mining is to explore new facts or generate novel hypothesis based on the published information. In earlier works, text mining approaches in biomedical research investigated to find out the answer of a specific problem in the related area. In [1], authors are interested in formulating a hypothesis about the contribution of some novel genes in a disease by analyzing their co expression with genes already proved to be involved in that particular disease. The method explores biomedical literatures and provides an interface for experimenting the association of different gene sequences in disease on the basis of the information published in those literatures. [2] Extracts the relation-ships among various biological entities from existing literatures. Their method provides a user interactive interface for querying iteratively and also ranks the extracted relationships by computing p-value of the assumed hypothesis which is claimed to be significant in drug discovery. ALIBABA [3] is an interactive tool that features several aspects including a graphical display of network consisting of all the extracted relationships of the queried entity with other entities cited in the publications. They have adopted both pattern matching and co-occurrence filtering method for exploring the relationships and provided confidence level to each of the edges such that user can easily prune the network up to their confidence threshold. iHOP [4] is another tool that integrates information about genes and makes a way for easy navigation from one gene information to an- other by hyper-linking particular corpus to the specific gene name. GoPubMed [5] is gene ontology based PubMed search which explores the literatures on the basis of the GO term related to the query and allows user to navigate the abstracts categorized on the relevant GO terms. BioIE retrieves abstracts from PubMed based on a keyword query and then identifies and extracts informative sentences using rule and pattern based methods [6]. EBIMed [7] retrieves a set of abstracts as a result of keyword search and then filters for sentences containing UniProtKB/Swiss-Prot proteins, GO terms, drugs and species. HubMed [9] is a web based biomedical literature search platform where users get an integrated interface equipped with a number of features for visualizing clusters of related articles and doing other supporting activities. The exploratory tool XplorMed [10] has been developed to analyse the result of any MEDLINE query. It is a pioneer to the automation of data mining and data organization in molecular biology.

The works described above are basically devoted to mining texts either for generating network centering the user query term or for supplying a set of supporting documents where user can get their relevant information. The hypothesis driven text mining tool is not available where user can prove or disprove any hypothesis based on text corpus. [11] provides an evolutionary technique to generate and test hypothesis on the basis of the number of documents supporting the hypothesis and the relationship strength as retrieved by Chilibot. But they havenot considered the type of relation the query terms have among them and the hypothesis is restricted to have only the concepts connected by AND, OR, NOT instead of any particular verb. BioGraph [12] provides an online resource and data mining method for the automated inference of functional hypotheses be-tween biomedical entities. The proposed method can rank the association among the entities (disease-gene association) and thus can nd out the most relevant entities to the source entity without any supervised learning. Another method builds a disease-gene network based on dependency parsing and support vector machine and then measures some statistical metrics to rank the genes in the network [13]. [14] introduces a tool CoPub Discovery that mines the literature for new relationships between genes, drugs, pathways and diseases. Several of the newly found relationships between compounds and cell proliferation have been validated using independent literature sources. Integrated Bio-Entity Net-work [15] is a recent framework that incorporates graph theoretic algorithms like Breadth- rst search with pruning (BFSP) and most probable path (MPP) algorithms to assign weight to the curated relationships and thus generate hy-pothesis from the existing literatures.

All these previous works are focused to build network consisting of biological entities from where users can get new knowledge about biological entities and their relationships. Most of them are either co-occurrence based or pattern based systems. In our method, we try to overcome the shortcomings of the previous works. Our method not only tests the hypothesis based on the statistical mea-sure but also builds a network comprising of the entities directly and indirectly involved with the referred entity in the hypothesis. On top of it, our approach attempts to explore the relationships by exploiting natural language processing technique which obviously increases its accuracy and precision. Thus it opens up the door to further research in the respective area. Besides this, our system pro-vides a graphical interface where user can flexibly put the hypothesis in natural English language and does not need to bother about the linguistic pattern. We also extend the tool for iterative hypothesis testing and graphical input system without textual interruption.

## 3 Method

### 3.1 Building Database

The very first step of the proposed method is to build a database consisting of all kinds of relationships among different biological entities cited in a paper. We extract all kinds of relationships from a set of papers and store them in a database. For building the database, we have to find out semantic meaning of each of the lines of that paper. If any line describes the positive relationship among the entities, database stores the information with positive value. On the contrary, if the relationship described in a line indicates any negative relation among the entities, database stores the information with negative value. Any relation cited multiple times in the paper is stored once in the database.

### 3.2 Natural Language Processing

The proposed system accepts the hypothesis to be tested in textual format. Initially we plan to work with simple sentence consisting of two biological entities and a verbal relationship among them. However, our system also works for any kind of complex or compound sentence describing relationship among the entities. To find out the supporting lines in different literatures our first step is to do some natural language processing such that we can extract the semantic meaning of those lines and thus be able to identify their actual role in supporting the given hypothesis. As for example, a paper contains a line \Protein A interacts Protein B". As two entities co-occur in that line, our method can easily extract the relationship and stores the information in the database. If any line describes the relation in the way that \It is not evident that the interaction of Protein A with Protein B really exists", we will categorize the line as having an opposing role to any relationship between Protein A and Protein B. This line will con-tribute negative relation among the entities and that relation will be stored in the database with negative indicator. On the contrary, the supporting lines will contribute positive factor and will be incorporated in the database with positive value. Similar case happens for processing input hypothesis. The input hypothesis is similarly analyzed to find out the query pattern among the cited entities in the hypothesis and then stored in the database.

### 3.3 p-value Calculation

Statistical analysis of the hypothesis to be tested will be able to provide user a quantitative measurement of the hypothesis. In the proposed method, the hypothesis is processed and normalized by natural language processing in a way discussed in the previous subsection. We are basically concerned about testing simple hypothesis declaring a relationship between biological entities. The normalization step tags the entity names uttered in the hypothesis and also finds out the synonyms or aliases along with the types of those entities from the existing databases. The semantic meaning of the relationship is also grabbed after stemming the verb and doing required processing of the hypothesis. Our method then executes database query to find out the role of the paper in supporting the given hypothesis. If it supports the hypothesis, the particular paper will be counted as observed data for the p-value calculation.

In the next step, we count the number of papers that supports the hypothesis to be tested. This number is treated as observed frequency whereas the number of papers expected to support the hypothesis is treated as expected frequency which can be chosen by the user. Then we do Chi-square test by computing the following formula:

$$\chi^2 = (o - e)^2 / e \quad (1)$$

to compare the observed data with data we would expect to prove or disprovespecific hypothesis. Chi-square is the sum of the squared difference between observed (o) and the expected (e) data (or the deviation, d), divided by the expected data in all possible categories. After calculating the value, we use the chi-square distribution table to determine the significance of the value. In our case, the degree of freedom is 1 (number of states of each variable - 1) as the number of states of each variable's (entity) existence in the paper can be of two types (might co-exist with another entity or not). Then we conclude the hypothesis either by accepting (p > .05) or rejecting the hypothesis (p < .05). According to chi-square distribution, p-value greater than .05 indicates that the deviation is small enough that chance alone accounts for it and, therefore, we accept the hypothesis to be true. On the contrary, the opposite case signifies that some factor other than chance is operating for the deviation to be so great in which case we reject the hypothesis. We choose chi-square distribution for measuring the quantitative value of the hypothesis as our method is based on the comparison of the observed result with the expected outcome. Using this method researcher can easily get an idea about their assumption from literature's perspective.

### 3.4 Biological Network Generation

In this approach, we not only evaluate hypothesis statistically but also provide a biological network surrounding the entities the researcher might be interested in. The biological network will help him to generate new assumption evolving the interested one and thus get resource for further investigation. The network is basically centered around the entities mentioned in the testing hypothesis. For example, the hypothesis to be tested is "Protein A interact Protein B". Our tool will testify the hypothesis and compute the p-value using chi-square test. In addition to this, it will extract all other entities which co-occur with the protein A or protein B or which make a bridge of connection between them through transitive relation. The network will be displayed graphically for the convenience of the user to get a comprehensive view of the inter relations among the entities. On top of this, user will be able to generate new hypothesis by selecting the nodes and edges in the graphical network which will then be converted to textual hypothesis and submitted as the previous one. Thus user can interactively generate new hypothesis and put them for testing flexibly without typing. The interface integrates all the functionalities such that user can be benefitted.

## 4 Algorithm

First of all we search each of the papers in the database to find biological entities. In 4.1, PAPER[i][j] points the j$^{th}$ line of i$^{th}$ paper. If biological entities have been found in any sentence, we try to extract semantic meaning of the sentence using the function FINDRELATION(PAPER[i][j]) defined in 4.1. This function processes each line of that particular paper by finding out the type of relation whether it is positive or negative. The function SAV ERELATION(Entity1; Entity2)stores all the positive relations with the value 1 and all the negative relations with the value 1. We traverse the whole paper and extract all kinds of relationships inside that paper. Then we process the hypothesis sentence using the way we do for finding out the existing relations in the papers. In the following step, we do a search in the stored data. If we find a path between u and v of the hypothesis in the data as a result of query execution in the database, it will return positive value. Positive return value means that the paper supports the hypothesis and increases the value of the counter. The function PGEN(counter; e) computes the p-value using Chi Square Distribution with the counter which is treated as observed frequency and e which is treated as expected frequency. If the return value of PGEN() is > :05, the hypothesis will be accepted, otherwise it will be rejected. The function SECONDARY NTWRK(u; v) in 4.1 actually constructs the graph where each node denotes biological entity and the edges represent the relationship among them. For secondary network generation, we emphasize on the entities which are directly or indirectly related to the entities mentioned in the hypothesis tested. The entities which are disconnected from the mentioned entities or which are found having no relation with those entities are not shown in the secondary network.

### 4.1 Program Code

```
programHypoTest (Output)
     n := Total number of papers;
     i := 1;
     j := 1;
     repeat
        l := Number of sentences in the ith line of the paper; repeat
          count NOB (Number of biological entities);
          if NOB >= 2
              Call Function FINDRELATION(PAPER[i][j]);
until j > 1
        Call Function SAVERELATION(A[i]);
until i > n
     h := Hypothesis sentence;
     Call Function FINDRELATION(h);
     i := 1;
     counter := 0;
     repeat
        Call Function DFS (u,v);
        If DFS(u,v)=1 then
        counter := counter + 1;
     until i > n
     Call Function PGEN (counter, e);
     if PGEN (counter, e) >= 0.05
        Accept h;
else
        Reject h;
  Call Function
  SECONDARYNTWRK(u,v); var
     i, j, n, counter, l: integer;
     u, v, h:= string;

FUNCTION FINDRELATION(s)
SNTWRK := 0;
Count := no. of complementary words;
IfCount%2==0 then
Flag := Positive;
else
Flag := Negative;
if Flag = Positive then
Relationship[Entitiy1][Entity2] := Relationship[Entitiy2][Entity1] := 1;
Call Function SAVERELATION(Entity1,Entity2);
    Else
        Relationship[Entitiy1][Entity2] := Relationship[Entitiy2][Entity1] := -1;

    FUNCTION SAVERELATION(Entity1,Entity2)
       SNTWRK[Entity1] := Entity2;
       SNTWRK[Entity1] := SNTWRK[Entity2];
       SNTWRK[Entity2] := Entity1;
       SNTWRK[Entity2] := SNTWRK[Entity1];
```

```
          SNTWRK[Entity1][Entity2] := v;

FUNCTIONSECONDARYNTWRK(u, v)
      Print u;
if SNTWRK[u] != 0 then
Print SNTWRK[u];
        else
            Print v;

    FUNCTION PGEN (counter, e)
         Result := (counter-e)2/e;
         PVALUE := P[Degreeoffreedom][Result];
         Return PVALUE;

End
```

## Results

We have tested 25 biological papers and extracted around 50 relations from them. In case of hypothesis testing, our achieved F-measure is 85%. The higher the value of F is, the accuracy will also be higher. We have stored a number of proven relationships in our database and then checked whether our system accepts any relation mentioned in the hypothesis which also exists in the database as true. We see that our system can accurately accept any hypothesis which is proven to be true and rejects any hypothesis which does not really exist. In Table 1, we have shown several examples of input hypothesis tested along with their calculated p-value by our system.

Table 1. Performance Analysis Table

| Input Hypothesis | o | e | p-value | Decision |
|---|---|---|---|---|
| Carvedilol not causes Weight Gain | 18/25 | 15/25 | 0.40 | Accepted |
| It is not evident that Carvedilol causes Weight Gain | 18/25 | 15/25 | 0.40 | Accepted |
| Bacillus Calmette-Gurin cures buruli ulcer | 14/25 | 12/25 | 0.42 | Accepted |
| Melanocortin4 receptor leads to severe weight gain | 13/25 | 14/25 | 0.82 | Accepted |
| It is not evident that MC4R causes diabetics | 2/25 | 25/35 | 0.001 | Rejected |

Now, we will explain some of the example given in the above table. Let us consider the hypothesis "Carvedilol not causes Weight Gain". After giving the input hypothesis, our expected outcome (e) was 15/25. But our observed result (e) was 18/25. By setting the values in formula (1), we get $x^2$= 0.6,by this value we can find the p-value through chi-square distributiontable, which is between 0.4 (approx.). As it is greater than 0.05, we have accepted the hypothesis.

## 6 User Interface

In order to make the tool available to the community of researchers, we develop a web based tool HypoTest shown in Fig. 1. First, user can submit his query hypothesis in textual format as in Fig 1(a). After submitting the query, the algorithm in the background will compute the p-value with respect to existing datasets and then show the result in the following page as shown in Fig 1(b). The tool shows not only the p-value of the given hypothesis, but also generates a biological network for the conveniences of the user. The tool also provides graphical input system for naive users as in Fig 1(c). That means user has two options for building the query hypothesis. He can submit the testing hypothesis either in textual format or in graphical format. The biological network broadens the scope of thinking of the researchers and he can build new hypothesis for testing and investigating. The user interactive interface of our system claims to be the first attempt in this area so far. It facilitates all classes of people to do their research in more flexible and cooperative way indeed.

# 7 CONCLUSION

The paper describes a statistical approach to compute the significance of the hypothesis submitted by user and thus provides a way to measure the feasibility

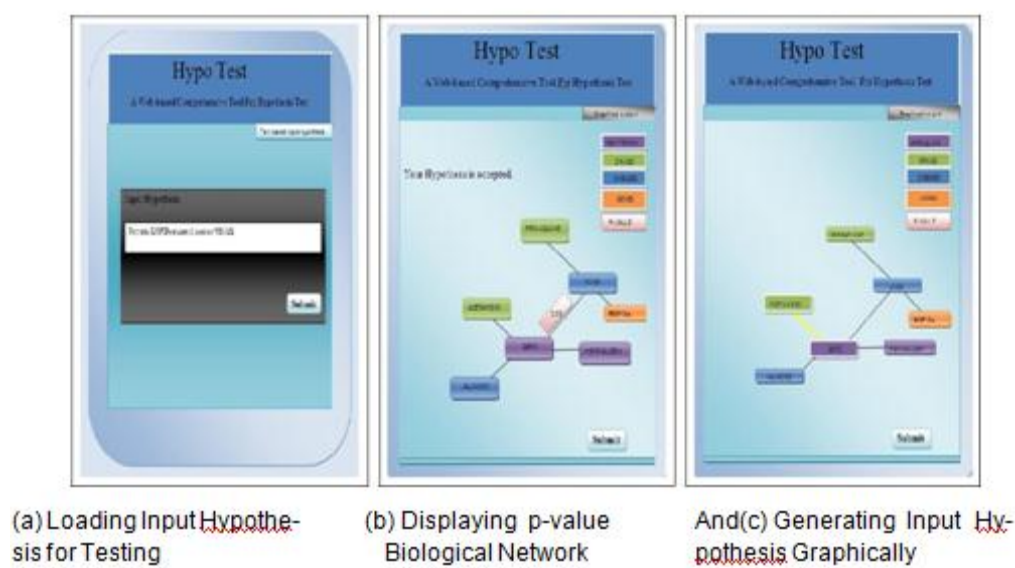

Fig. 1. Interface

Ofthe hypothesis quantitatively. Moreover, the method comes up with an interface for the greater conveniences of the user where he will also be able to put the query graphically and visualize the biological network surrounding the entities of his interests. As a result, he will get scope to think further and generate new hypothesis from the network and iterate the process until he gets satisfactory outcomes. Our tool claims to be the first initiative in the area of significance measurement of the hypothesis quantitatively. The graphical input system and biological network generation bring flexibility to the researcher and lead to the new way of thinking. In future, we plan to work with complex hypothesis generation graphically and their significance computation in the context of existing literatures.